\documentclass[a4paper]{article}

\usepackage{amsmath}
\usepackage{amssymb}
\usepackage{amsfonts}
\usepackage{mathrsfs}
\usepackage{authblk}
\usepackage{graphicx}
\usepackage[pdftex]{hyperref}

\newcommand{\red}{}

\newcommand{\ph}{\circ}
\newcommand{\rng}{\text{rng}}
\newcommand{\eff}{\text{eff}}

\newcommand{\EP}{E_\text{P}}

\newcommand{\hhtt}{h^\text{TT}}
\newcommand{\pp}{\boldsymbol{p}}
\newcommand{\qq}{\boldsymbol{q}}
\newcommand{\kk}{{\boldsymbol{k}}}
\newcommand{\rr}{\boldsymbol{r}}

\newcommand{\zz}{\boldsymbol{z}}
\newcommand{\hc}{\text{H.c.}}
\newcommand{\dd}{\text{d}}
\newcommand{\TT}{\text{TT}}
\newcommand{\LL}{{\mathcal L}}
\newcommand{\bra}[1]{\langle#1|}

\newcommand{\ket}[1]{|#1\rangle}
\newcommand{\ppp}{\\[3pt]}
\newcommand{\nppp}{\nonumber\\[3pt]}
\newcommand{\pppp}{\\[6pt]}

\begin{document}

\title{Bremsstrahlung of Light through Spontaneous Emission of Gravitational Waves}

\author{Charles H.-T. Wang}

\author{Melania Mieczkowska}

\affil{Department of Physics, University of Aberdeen, King's College, Aberdeen AB24 3UE, United Kingdom}

\date{11 May 2021}

\maketitle


\begin{center}
Published in the Special Issue of 
\href{https://www.mdpi.com/journal/symmetry}{Symmetry}: \href{https://www.mdpi.com/journal/symmetry/special_issues/New_Frontiers_Quantum_Gravity}
{New Frontiers in Quantum Gravity}
\end{center}

\begin{abstract}
Zero-point fluctuations are a universal consequence of quantum theory. Vacuum fluctuations of electromagnetic field have provided crucial evidence and guidance for QED as a successful quantum field theory  with a defining gauge symmetry through the Lamb shift, Casimir effect, and spontaneous emission. In an accelerated frame, the thermalisation of the zero-point electromagnetic field gives rise to the Unruh effect linked to the Hawking effect of a black hole via the equivalence principle.  This principle is the basis of general covariance, the symmetry of general relativity as the classical theory of gravity. If quantum gravity exists, the quantum vacuum fluctuations of the gravitational field should also lead to the quantum decoherence and dissertation of general forms of energy and matter. Here we present a novel theoretical effect involving the spontaneous emission of soft gravitons by photons as they bend around a heavy mass and discuss its observational prospects. Our analytic and numerical investigations suggest that the gravitational bending of starlight predicted by classical general relativity should also be accompanied by the emission of gravitational waves. This in turn redshifts the light causing a loss of its energy somewhat analogous to the bremsstrahlung of electrons by a heavier charged particle.  It is suggested that this new effect may be important for a combined astronomical source of intense gravity and high-frequency radiation such as X-ray binaries and that the proposed LISA mission may be potentially sensitive to the resulting sub-Hz stochastic gravitational waves.
\end{abstract}

\vspace{10pt}
{\bf DOI}:
\href{https://doi.org/10.3390/sym13050852}
{10.3390/sym13050852}


\section{Introduction}

The quantum vacuum is a remarkable consequence of the quantum field theory (QFT). To be sure, the quantum electrodynamics (QED) as the first successful QFT has received crucial guidance and support through its quantum vacuum effects including the Lamb shift, Casimir effect, and spontaneous emission.

Although the physical reality of the quantum vacuum seems to contradict the void classical vacuum, it in fact forges essential links between classical and quantum dynamics. The general agreement between the classical emission rate and quantum spontaneous emission rate of electromagnetic (EM) dipole radiations have been well-known at atomic scales (see e.g., \cite{Jaynes1963}). Such an agreement is also clear in the classical cyclotron radiation and the quantum spontaneous emission of the Landau levels \cite{Gregori2021}, in the context of detecting Unruh radiation as a quantum vacuum effect in non-inertial frames \cite{Crowley2012}.

\newpage

At present, a fully quantised theory of gravity is still to be reached (for some recent developments, see e.g., ~\cite{Wang2020, Wang2018, Veraguth2017} and references therein). Nevertheless, the effective QFT for linearised general relativity is expected to yield satisfactory physical descriptions at energies sufficiently lower than the Planck scale
\cite{Burgess2004, Oniga2016a, Oniga2016b}.
Indeed, the spontaneous emission rate of gravitons for a nonrelativistic bound system due to the zero-point fluctuations of spacetime in linearised quantum gravity has been recently shown \cite{Oniga2017b, Quinones2017} to agree with the quadrupole formula of gravitational wave radiation in general relativity \cite{Misner1973}.  The preservation of the local translational symmetry of linearised gravity is crucial in the theoretical steps of establishing this agreement through the gauge invariant Dirac quantisation technique \cite{Oniga2016a}.

Based on this development, the next challenge would be the spontaneous emission of gravitons from a relativistic and unbound system, which we will address in this paper. The possible gravitational radiation by photons has long been a subject of interest and has been considered by many researchers with various approaches
\cite{Oniga2017a, Baker2012, Raffelt1999, Papini1989, Campo1988, Gould1985, Ford1982, Grishchuk1977, Grishchuk1973}.
The obtained size of the effect has generally been quite small.

We therefore seek an amplified effect in the astronomical context involving the deflection of starlight by a celestial body or distribution of mass. We show that soft gravitons are spontaneously emitted resulting in scattering modes of incident photons to decay into lower energy scattering modes in the fashion of the bremsstrahlung of electrons by ions~\cite{Lebed2010, Breuer2001, Karapetyan1978, Bethe1934}. Our preliminary estimates of such effects suggest they may be important for high frequency photons deflected by a compact heavy mass.

Under weak gravity, the polarisations of light subject to gravitational bending are expected to be negligible. Therefore, as a first approximation, the effect of spin of photon is neglected similar to neglecting spins in standard descriptions of the bremsstrahlung of~electrons.

Throughout this paper, we adopt units in which the speed of light $c$ equals unity, unless otherwise stated. We also write $\log_{10}=\log$ and use Greek indices $\mu,\nu\ldots=0,1,2,3$ and Latin indices $i,j\ldots=1,2,3$ for spacetime $(t,x,y,z)$ and spatial $(x,y,z)$ coordinates,~respectively.

\section{Light Modelled as Massless Scalar Field in a Weak Central Gravitational Field}

\begin{figure}
\begin{center}
\includegraphics[width=0.9\linewidth]{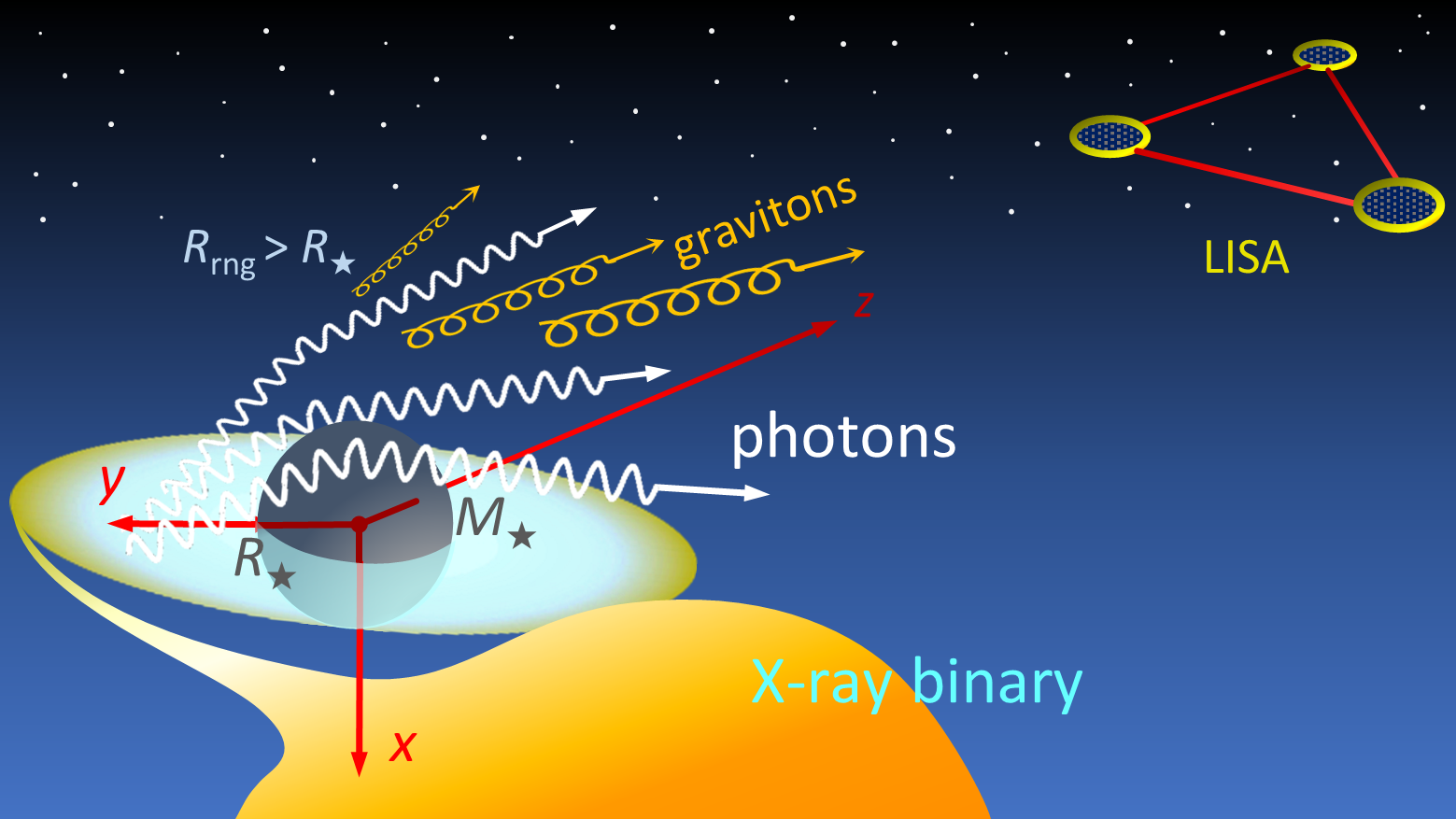}
\caption{{\red An illustration of the key physical and geometrical features of an astronomical configuration for the gravitational bremsstrahlung of light involving an X-ray binary and a possible detection concept with LISA. Here the mass of the compact object is assumed to be dominant for simplicity.}}
\label{fig.1}
\end{center}
\end{figure}

As alluded to in the introduction section, in what follows, we shall model photons as massless scalar particles with a linearised metric
\begin{eqnarray}
g_{\mu\nu}
&=&
\eta_{\mu\nu}+h_{\mu\nu}
\label{getah}
\end{eqnarray}
where
$\eta_{\mu\nu}=$ diag$(-1,+1,+1,+1)$ is the Minkowski metric and
$h_{\mu\nu}$ is the metric perturbation arising from a spherical gravitational source with mass $M_\star$ so that
\begin{eqnarray}
h_{00}
&=&
h_{11}
=
h_{22}
=
h_{33}
=
-2\Phi
\label{h00}
\end{eqnarray}
in terms of the Newtonian potential \cite{Misner1973}
\begin{eqnarray}
\Phi
&=&
-\frac{G M_\star}{r}
\label{ntpol}
\end{eqnarray}
with $r = \sqrt{x^2+y^2+z^2}$ and the gravitational constant $G$. Fig.~\ref{fig.1} illustrates the schematic physical and geometrical configurations under consideration.
Treating the above source as a gravitational lens, its effective refractive index is given approximately by $n=1-2\Phi$ \cite{Meylan2006}.
This gives rise to the approximate dispersion relation
\begin{eqnarray}
\omega
&=&
(1+2\Phi)p
\label{omp}
\end{eqnarray}
for a real massless scalar field $\phi$ having a frequency $\omega$ and wave vector $\pp$ with wave number $p = |\pp|$. We will continue to denote wave vectors associated with the scalar field by $\pp$ and $\qq$ and wave vectors associated with the gravitons by $\kk$, unless otherwise stated.

To capture the salient physical effects carried by the light frequency and to simplify our technical derivations, we consider the wave number of the scalar to peak around some fixed value $p_\ph$. Then, it follows from Eq.~\eqref{omp} to leading contributions, that
\begin{eqnarray}
\omega^2
&=&
p^2 + v(r)
\end{eqnarray}
where
\begin{eqnarray}
v(r)
&=&
-4p_\ph^2\frac{G M_\star}{r} .
\label{vPhi}
\end{eqnarray}

Using Eq.~\eqref{vPhi}, the Lagrangian density of the scalar field
\begin{eqnarray*}
\LL
&=&
-\frac12\,\sqrt{-g}\,
g_{\mu\nu}\nabla^{\mu}\phi\nabla^{\nu}\phi
\end{eqnarray*}
reduces to
\begin{eqnarray}
\LL
&=&
-
\frac12
\eta^{\mu\nu}\phi_{,\mu}\phi_{,\nu}
-\frac12 v(r) \phi^2
\label{Lmsclreff}
\end{eqnarray}
with $v(r)$ as the effective external scalar potential \cite{Milton2011, Milton2016}.


The stress-energy tensor of the scalar field is given by
\begin{eqnarray}
T^{\mu\nu}
&=&
\frac12
(\eta^{\mu\alpha}\eta^{\nu\beta}+\eta^{\mu\beta}\eta^{\nu\alpha}-\eta^{\mu\nu}\eta^{\alpha\beta})
\phi_{,\alpha}\phi_{,\beta}
-
\frac12\eta^{\mu\nu} v \phi^2 .
\label{ser}
\end{eqnarray}

The field equation follows as
\begin{eqnarray}
\eta^{\mu\nu}
\partial_\mu\partial_\nu\phi
- v \phi
&=&
0 .
\label{seq}
\end{eqnarray}

To solve this field equation, one naturally invokes the separable ansatz
\begin{eqnarray}
\phi(\rr,t)
=
\psi(\rr)e^{-i\omega t}
\label{Phi}
\end{eqnarray}
so that the general solutions are the real parts of the linear combinations of Eq.~\eqref{Phi}.
Substituting the above ansatz into Eq.~\eqref{seq}, we see that this field equation is equivalent to
\begin{eqnarray}
-\nabla^2\psi + v\psi
=
\omega^2\psi
\label{seqc}
\end{eqnarray}
taking the form of a time-independent Schr\"odinger equation.


It follows that the solutions to the field equation \eqref{seq} representing the deflection of light with an incident wave vector $\pp$ and frequency $\omega=p$ can be obtained from the solutions of the Schr\"odinger equation \eqref{seqc} describing a scattering problem involving a Coulomb-type, i.e. $1/r$, central potential as the ``scattering wave functions'' of the form
\begin{eqnarray}
\psi_{\pp}(\rr)
&=&
\frac{1}{\rho}
\sum_{l=0}^{\infty}\sum_{m=-l}^{l}
4\pi\, i^l w_l(\eta,\rho) Y_l^m(\hat{\rr}) Y_l^{m*}(\hat{\pp})
\label{psik}
\end{eqnarray}
where $\hat{\pp}=\pp/p$, $\rr=(x,y,z)$, $Y_l^m(\rr)$ are spherical harmonics, and
$w_l(\eta,\rho)$ are the Coulomb wave functions satisfy the wave equation (see e.g. Ref.~\cite{Wong2004})
\begin{eqnarray}
\Big\{
\frac{\dd^2}{\dd\rho^2}
+
\Big[
1-\frac{2\eta}{\rho}-\frac{l(l+1)}{\rho^2}
\Big]
\Big\}
w_l(\eta,\rho)
&=&
0
\label{cweq}
\end{eqnarray}
using the dimensionless variable $\rho = p\, r$ and dimensionless parameter
$\eta = -{\nu}/{p}$
with
\begin{eqnarray}
\nu
&=&
2\, G M_\star p_\ph^2 .
\label{nu}
\end{eqnarray}

By virtue of the orthogonality of $w_l(\eta,\rho)$, we can choose the normalisation of $w_l(\eta,\rho)$ so that $\psi_{\pp}(\rr)$ satisfy the following orthonormality
\begin{eqnarray}
\int\dd^3r\, \psi_{\pp'}^*(\rr) \psi_{\pp}(\rr)
&=&
\delta(\pp-\pp') ,
\label{orthk}
\ppp
\int{\dd^3p}\, \psi_{\pp}^*(\rr') \psi_{\pp}(\rr)
&=&
\delta(\rr-\rr') .
\label{orthr}
\end{eqnarray}
%


It is useful to introduce the ``momentum representation'' scattering wave functions \cite{Guth1951} $\psi_{\qq}(\pp)$ of the above ``position representation'' scattering wave functions $\psi_{\qq}(\rr)$ given by
\begin{eqnarray}
\psi_{\qq}(\pp)
&=&
\frac{1}{(2\pi)^{3/2}}\int\dd^3r\,e^{-i\pp\cdot\rr}\psi_{\qq}(\rr) ,
\label{r2p}
\ppp
\psi_{\qq}(\rr)
&=&
\frac{1}{(2\pi)^{3/2}}\int\dd^3p\,e^{i\pp\cdot\rr}\psi_{\qq}(\pp) .
\label{p2r}
\end{eqnarray}
%



The orthogonality of $\psi_{\qq}(\pp)$  follows immediately from Eqs.~\eqref{orthk}, \eqref{orthr}, \eqref{p2r} and \eqref{r2p} to be
\begin{eqnarray}
\int\dd^3q\, \psi_{\qq}^*(\pp') \psi_{\qq}(\pp)
&=&
\delta(\pp-\pp') ,
\label{orthp}
\ppp
\int\dd^3q\, \psi_{\pp'}^*(\qq) \psi_{\pp}(\qq)
&=&
\delta(\pp-\pp') .
\label{orthq}
\end{eqnarray}

For a weak interaction with the central potential where $|\eta| \ll 1$, the first order Born approximation yields
\begin{eqnarray}
\psi_{\pp}(\qq)
&=&
\delta(\pp-\qq) + \frac{\nu}{\pi^2(p^2-q^2)|\pp-\qq|^2}
\label{psikkpp}
\end{eqnarray}

Note that the first term of Eq.~\eqref{psikkpp} corresponds to the first term of Eq.~\eqref{psikkrr} which represents the incident (asymptotically) free particle. This term does not contribute to Eq.~\eqref{tauk_ij} under Markov approximation of the gravitational master equation as discussed in Ref.~\cite{Oniga2016b}.

\newpage

The corresponding asymptotic scattering wave function of the position $\rr$ is given by
\begin{eqnarray}
\psi_{\pp}(\rr)
&=&
\frac{1}{(2\pi)^{3/2}}
\Big[\,
e^{i\pp\cdot\rr} + f(\theta) \frac{e^{i p r}}{r}\,
\Big]
\label{psikkrr}
\end{eqnarray}
with the scattering amplitude
\begin{eqnarray}
f(\theta)
=
\frac{\nu}{2 p^2 \sin^2(\theta/2)}
\label{frr}
\end{eqnarray}
where
$\theta=\measuredangle(\rr,\pp)$
is the scattering angle.

Just as with the Coulomb potential, so does the infinitely long range of the Newtonian potential imply a divergent total scattering cross-section. However, to account for the realistic limited dominance of this potential due to other influences beyond a  range distance $R_\rng=1/\epsilon$ which can be conveniently incorporated by modifying the Newtonian potential with an additional exponential-decay factor of $e^{-\epsilon r}$ as a long-range regularisation. On the other hand, the finite extension with a radius $R_\star=1/\delta$ of the gravity source means the need for a compensating short-range potential within this radius.

These considerations lead to the following phenomenological Yukawa regularisation
\begin{eqnarray}
\frac1{r}
\to
\frac{e^{-\epsilon r}}{r}
-
\frac{e^{-\delta r}}{r}
\end{eqnarray}
with $\epsilon$ and $\delta$ as long and short range regularisation parameters respectively.

Accordingly, we find the regularised scattering wave function to be
\begin{eqnarray}
\psi_{\qq}(\pp)
&=&
\delta(\pp-\qq) + \frac{\nu}{\pi^2[p^2-(q+i\epsilon)^2][|\pp-\qq|^2+\epsilon^2]}
\nppp&&
-
\frac{\nu}{\pi^2[p^2-(q+i\delta)^2][|\pp-\qq|^2+\delta^2]}
\label{psikkppy}
\end{eqnarray}
with $p_\ph \gg \delta > \epsilon$.


\section{Quantisation of the Scalar Field in the Regularised Potential}
\label{sec:qsf}

Using scattering wave function $\psi_{\pp}(\rr)$ derived in the preceding section, we can now perform the so-called second quantisation of the scalar field $\phi$ into a quantum field operator in the Heisenberg picture as follows
\begin{eqnarray}
\phi(\rr,t)
&=&
\int\dd^3p\,
\sqrt{\frac{\hbar}{2\omega_{p}}}\;
\Big[
a_{\pp}\psi_{\pp}(\rr)e^{-i\omega_{\pp} t}
+
a_{\pp}^\dag\psi_{\pp}^*(\rr)e^{i\omega_{\pp} t}
\Big]
\label{phiex}
\end{eqnarray}
where $\hbar$ is the reduced Planck constant and the creation and annihilation operators $a_{\pp}$ and $a_{\pp}^\dag$ satisfy the standard nontrivial canonical commutation relation
\begin{eqnarray}
\big[\,
a_{\pp}, a_{\pp'}^\dag
\big]
&=&
\delta(\pp-\pp') .
\label{comma}
\end{eqnarray}

The associated field momentum is given by $\pi= \partial\phi/\partial t = \dot{\phi}$, which satisfies the equal time field commutation relation
\begin{eqnarray*}
[\phi(\rr,t),\pi(\rr',t)]
&=&
i\hbar\,\delta(\rr-\rr')
\end{eqnarray*}
following from Eqs.~\eqref{orthr} and \eqref{comma}.

Substituting Eq.~\eqref{p2r} into Eq.~\eqref{phiex}, we can write $\phi$ in terms of the momentum representations of the scattering wave functions as follows
\begin{eqnarray}
\phi(\rr,t)
&=&
\int\dd^3q\,\dd^3p\,
\sqrt{\frac{\hbar}{2(2\pi)^3{q}}}\;
\Big[
a_{\qq}\psi_{\qq}(\pp)e^{i\pp\cdot\rr}e^{-i{q} t}
+
a_{\qq}^\dag\psi_{\qq}^*(\pp)e^{-i\pp\cdot\rr}e^{i{q} t}
\Big] .
\label{phiex2}
\end{eqnarray}

The coupling of $\phi$ to the metric fluctuations due to low energy quantum gravity in addition to the metric perturbation Eq.~\eqref{h00} due to the lensing mass $M_\star$ is through the transverse-traceless (TT) part of its stress-energy tensor Eq.~\eqref{ser} to be $\tau_{ij}:=T_{ij}^\TT$ \cite{Oniga2016a} given by
\begin{eqnarray}
\tau_{ij}(\rr, t)
&=&
P_{ijkl}\phi_{,k}(\rr, t)\phi_{,l}(\rr, t) ,
\label{tau0_ij}
\end{eqnarray}
with the Fourier transform
\begin{eqnarray}
\tau_{ij}(\kk, t)
&=&
\int\dd^3r \,
\tau_{ij}(\rr, t)
\,e^{-i \kk\cdot\rr}
\nppp
&=&
\int\dd^3r \,
e^{-i \kk\cdot\rr}
P_{ijkl}(\kk)\phi_{,k}(\rr, t)\phi_{,l}(\rr, t)
\label{tau_ij}
\end{eqnarray}
where $P_{ijkl}$ is the TT projection operator \cite{Oniga2016a, Flanagan2005}.

Using Eq.~\eqref{phiex2} and applying normal orders and neglecting $a_{\pp}a_{\qq}$ and $a_{\pp}^\dag a_{\qq}^\dag$ terms through the rotating wave approximation \cite{Breuer2002}, we see that Eq.~\eqref{tau_ij} becomes
\begin{eqnarray}
\tau_{ij}(\kk, t)
&=&
\frac{\hbar}{2}\,P_{ijkl}(\kk)
\int\dd^3p\,\dd^3p'\,\dd^3p''\,
\frac{p_k p_l}{\sqrt{{p'}{p''}}}
\nppp&&
\Big[
a_{\pp''}^\dag a_{\pp'}
\psi_{\pp'}(\pp)\psi_{\pp''}^*(\pp-\kk)
e^{-i({p'}-{p''}) t}
+
a_{\pp'}^\dag a_{\pp''}
\psi_{\pp'}^*(\pp)\psi_{\pp''}(\pp+\kk)
e^{i({p'}-{p''}) t}
\Big] .
\label{tauk_ij}
\end{eqnarray}

From Eq.~\eqref{tauk_ij} we see that
\begin{eqnarray}
\tau_{ij}(-\kk, t)
&=&
\tau_{ij}^\dag(\kk, t)
\label{aex}
\end{eqnarray}
as a useful property for later derivations.

\section{Coupling to the Gravitational Quantum Vacuum}
\label{sec:cgq}

To induce the spontaneous emission of the photon lensed by the regularised Newtonian potential, we now include an additional TT gravitational wave-like metric perturbation $\hhtt_{\mu\nu}$ into Eq.~\eqref{getah} which carries spacetime fluctuations at zero temperature \cite{Oniga2016a}. The photon is assumed to travel a sufficiently long distance and time past the gravitational lens (see justifications below) so that we can neglect any memory effects in its statistical interactions with spacetime fluctuations. Additionally, we consider the energy scale to be low enough for the self interaction of the photon to be negligible. This leads to the Markov quantum master equation
\begin{eqnarray}
\dot\rho(t)
&=&
-
\frac{\kappa}{\hbar}
\int\! \frac{\dd^3 k}{2(2\pi)^3k}
\int_{0}^{\infty} \dd s\,
e^{-i k s}
[
\tau^\dag_{ij} (\kk,t),\,
\tau_{ij}(\kk,t-s) \rho(t)
]
+ \hc
\label{maseqnmkc}
\end{eqnarray}
in the interaction picture, where $\kappa = 8 \pi G$, $\tau_{ij}(\kk,t)$ is given by Eq.~\eqref{tauk_ij}, and $\hc$ denotes the Hermitian conjugate of a previous term,
for the reduced density operator, i.e. density matrix, $\rho(t)$ of the photon by averaging, i.e. tracing, out the degrees of freedom in the noisy gravitational environment \cite{Oniga2016a, Oniga2017b}.

It is convenient to express Eq.~\eqref{tauk_ij} in the form
\begin{eqnarray}
\tau_{ij}(\kk,t)
=
\int\dd^3p'\dd^3p''\,
\Big[
\tau_{ij}(\kk,\pp',\pp'')
e^{-i({p'}-{p''}) t}
+
\tau_{ij}^\dag(-\kk,\pp',\pp'')
e^{i({p'}-{p''}) t}
\Big]
\label{tkksc}
\end{eqnarray}
in terms of
\begin{eqnarray}
&&\hspace{-20pt}
\tau_{ij}(\kk,\pp',\pp'')
=
\frac{\hbar}{2}\,P_{ijkl}(\kk)
\int\dd^3p\,
\frac{p_k p_l}{\sqrt{{p'}{p''}}}\,
a_{\pp''}^\dag a_{\pp'}
\psi_{\pp'}(\pp)\psi_{\pp''}^*(\pp-\kk)
\label{stij1kk}
\end{eqnarray}

From Eqs.~\eqref{tkksc} and \eqref{aex} we also have
\begin{eqnarray}
\tau_{ij}^\dag(\kk,t)
=
\int\dd^3q'\dd^3q''\,
\Big[
\tau_{ij}(-\kk,\qq',\qq'')
e^{-i({q'}-{q''}) t}
+
\tau_{ij}^\dag(\kk,\qq',\qq'')
e^{i({q'}-{q''}) t}
\Big]
\label{tkkscd}
\end{eqnarray}

Substituting Eq.~\eqref{tkksc} into Eq.~\eqref{maseqnmkc} we have
\begin{eqnarray}
\dot\rho(t)
&=&
-
\frac{\kappa}{\hbar}
\int\! \frac{\dd^3 k\,\dd^3 p'\,\dd^3 p''}{2(2\pi)^3k}
\int_{0}^{\infty} \dd s\,
e^{-i (k-{p'}+{p''})s}
e^{-i({p'}-{p''}) t}
\big\{
[
\tau^\dag_{ij} (\kk,t),\,
\tau_{ij}(\kk,\pp',\pp'')
\rho(t)
]
+ \hc
\big\}
\nppp&&
-
\frac{\kappa}{\hbar}
\int\! \frac{\dd^3 k\,\dd^3 p'\,\dd^3 p''}{2(2\pi)^3k}
\int_{0}^{\infty} \dd s\,
e^{-i (k+{p'}-{p''}) s}
e^{i({p'}-{p''}) t}
\big\{
[
\tau^\dag_{ij} (\kk,t),\,
\tau_{ij}^\dag(-\kk,\pp',\pp'')
\rho(t)
]
+ \hc
\big\} .
\label{maseqnmkd}
\end{eqnarray}

We then apply the Sokhotski-Plemelj theorem
\begin{equation}
\int_0^\infty \dd s\, e^{-i \epsilon s} = \pi \delta(\epsilon) - i \mathbf{P} \frac{1}{\epsilon}
\label{sp2}
\end{equation}
where $\mathbf{P}$ is the Cauchy principal value. Since this Cauchy principal value contributes to a renormalised energy which can be absorbed in physical energies \cite{Breuer2002}, we can neglect it.

The remaining part of Eq.~\eqref{maseqnmkd} on account of Eq.~\eqref{tkkscd} is
\begin{eqnarray}
\dot\rho(t)
&=&
-
\frac{\pi\kappa}{\hbar}
\int \frac{\dd^3 k\,\dd^3 p'\,\dd^3 p''\,\dd^3 q'\,\dd^3 q''}{2(2\pi)^3k}
\delta(k-({p'}-{p''}))
e^{-i[k+({q'}-{q''})] t}
\nppp&&
\;\;\times\;
\big\{
[
\tau_{ij}(-\kk,\qq',\qq'')
,\,
\tau_{ij}(\kk,\pp',\pp'')
\rho(t)
]
+ \hc
\big\}
\nppp&&
-
\frac{\pi\kappa}{\hbar}
\int \frac{\dd^3 k\,\dd^3 p'\,\dd^3 p''\,\dd^3 q'\,\dd^3 q''}{2(2\pi)^3k}
\delta(k-({p'}-{p''}))
e^{-i[k-({q'}-{q''})] t}
\nppp&&
\;\;\times\;
\big\{
[
\tau_{ij}^\dag(\kk,\qq',\qq'')
,\,
\tau_{ij}(\kk,\pp',\pp'')
\rho(t)
]
+ \hc
\big\}
\nppp&&
-
\frac{\pi\kappa}{\hbar}
\int \frac{\dd^3 k\,\dd^3 p'\,\dd^3 p''\,\dd^3 q'\,\dd^3 q''}{2(2\pi)^3k}
\delta(k+({p'}-{p''}))
e^{-i[k+({q'}-{q''})] t}
\nppp&&
\;\;\times\;
\big\{
[
\tau_{ij}(-\kk,\qq',\qq'')
,\,
\tau_{ij}^\dag(-\kk,\pp',\pp'')
\rho(t)
]
+ \hc
\big\}
\nppp&&
-
\frac{\pi\kappa}{\hbar}
\int \frac{\dd^3 k\,\dd^3 p'\,\dd^3 p''\,\dd^3 q'\,\dd^3 q''}{2(2\pi)^3k}
\delta(k+({p'}-{p''}))
e^{-i[k-({q'}-{q''})]t}
\nppp&&
\;\;\times\;
\big\{
[
\tau_{ij}^\dag(\kk,\qq',\qq'')
,\,
\tau_{ij}^\dag(-\kk,\pp',\pp'')
\rho(t)
]
+ \hc
\big\} .
\label{maseqnmkf3}
\end{eqnarray}
%


Due to the weakness of interactions, the density matrix will evolve only slightly over time so that
\begin{eqnarray}
\rho(t)=\rho_0+\Delta\rho(t)
\end{eqnarray}
where $\rho_0=\rho(t=0)$ is the initial density matrix and the components of $\Delta\rho(t)$ are small compared to the components of $\rho_0$.

Then Eq.~\eqref{maseqnmkf3} yields
\begin{eqnarray}
\Delta\rho(t)
&=&
-
\frac{\pi\kappa}{\hbar}
\int \frac{\dd^3 k\,\dd^3 p'\,\dd^3 p''\,\dd^3 q'\,\dd^3 q''}{2(2\pi)^3k}
\delta(k-({p'}-{p''}))
\int_0^t\dd s\,e^{-i[k+({q'}-{q''})] s}
\nppp&&
\;\;\times\;
\Big\{
[
\tau_{ij}(-\kk,\qq',\qq'')
,\,
\tau_{ij}(\kk,\pp',\pp'')
\rho_0
]
+ \hc
\Big\}
\nppp&&
-
\frac{\pi\kappa}{\hbar}
\int \frac{\dd^3 k\,\dd^3 p'\,\dd^3 p''\,\dd^3 q'\,\dd^3 q''}{2(2\pi)^3k}
\delta(k-({p'}-{p''}))
\int_0^t\dd s\,e^{-i[k-({q'}-{q''})] s}
\nppp&&
\;\;\times\;
\Big\{
[
\tau_{ij}^\dag(\kk,\qq',\qq'')
,\,
\tau_{ij}(\kk,\pp',\pp'')
\rho_0
]
+ \hc
\Big\}
\nppp&&
-
\frac{\pi\kappa}{\hbar}
\int \frac{\dd^3 k\,\dd^3 p'\,\dd^3 p''\,\dd^3 q'\,\dd^3 q''}{2(2\pi)^3k}
\delta(k+({p'}-{p''}))
\int_0^t\dd s\,e^{-i[k+({q'}-{q''})] s}
\nppp&&
\;\;\times\;
\Big\{
[
\tau_{ij}(-\kk,\qq',\qq'')
,\,
\tau_{ij}^\dag(-\kk,\pp',\pp'')
\rho_0
]
+ \hc
\Big\}
\nppp&&
-
\frac{\pi\kappa}{\hbar}
\int \frac{\dd^3 k\,\dd^3 p'\,\dd^3 p''\,\dd^3 q'\,\dd^3 q''}{2(2\pi)^3k}
\delta(k+({p'}-{p''}))
\int_0^t\dd s\,e^{-i[k-({q'}-{q''})]s}
\nppp&&
\;\;\times\;
\Big\{
[
\tau_{ij}^\dag(\kk,\qq',\qq'')
,\,
\tau_{ij}^\dag(-\kk,\pp',\pp'')
\rho_0
]
+ \hc
\Big\} .
\label{maseqnmkf5}
\end{eqnarray}

\newpage

For $t\to\infty$, physically corresponding to $t$ greater than the effective interaction time of the system, we can again apply the Sokhotski-Plemelj theorem Eq.~\eqref{sp2} to Eq.~\eqref{maseqnmkf5} and neglect the Cauchy principal value terms to obtain
\begin{eqnarray}
\Delta\rho
&:=&
\Delta\rho(t\to\infty)
\nppp
&=&
-
\frac{\pi\kappa}{\hbar}
\int \frac{\dd^3 k\,\dd^3 p'\,\dd^3 p''\,\dd^3 q'\,\dd^3 q''}{2(2\pi)^3k}
\,\delta(k+({q'}-{q''}))
\,\delta(k-({p'}-{p''}))
\nppp&&
\;\;\times\;
\Big\{
[
\tau_{ij}(-\kk,\qq',\qq'')
,\,
\tau_{ij}(\kk,\pp',\pp'')
\rho_0
]
+ \hc
\Big\}
\nppp&&
-
\frac{\pi\kappa}{\hbar}
\int \frac{\dd^3 k\,\dd^3 p'\,\dd^3 p''\,\dd^3 q'\,\dd^3 q''}{2(2\pi)^3k}
\,\delta(k-({q'}-{q''}))
\,\delta(k-({p'}-{p''}))
\nppp&&
\;\;\times\;
\Big\{
[
\tau_{ij}^\dag(\kk,\qq',\qq'')
,\,
\tau_{ij}(\kk,\pp',\pp'')
\rho_0
]
+ \hc
\Big\}
\nppp&&
-
\frac{\pi\kappa}{\hbar}
\int \frac{\dd^3 k\,\dd^3 p'\,\dd^3 p''\,\dd^3 q'\,\dd^3 q''}{2(2\pi)^3k}
\,\delta(k+({q'}-{q''}))
\,\delta(k+({p'}-{p''}))
\nppp&&
\;\;\times\;
\Big\{
[
\tau_{ij}(-\kk,\qq',\qq'')
,\,
\tau_{ij}^\dag(-\kk,\pp',\pp'')
\rho_0
]
+ \hc
\Big\}
\nppp&&
-
\frac{\pi\kappa}{\hbar}
\int \frac{\dd^3 k\,\dd^3 p'\,\dd^3 p''\,\dd^3 q'\,\dd^3 q''}{2(2\pi)^3k}
\,\delta(k-({q'}-{q''}))
\,\delta(k+({p'}-{p''}))
\nppp&&
\;\;\times\;
\Big\{
[
\tau_{ij}^\dag(\kk,\qq',\qq'')
,\,
\tau_{ij}^\dag(-\kk,\pp',\pp'')
\rho_0
]
+ \hc
\Big\} .
\label{maseqnmkf6}
\end{eqnarray}

More precisely, the limit $t\to\infty$, means $t$ exceeds the effective interaction duration between the photon and the lensing mass with an effective force range $\sim 1/\epsilon$, and this is satisfied if the photon is measured at a large distance with $t \gg 1/\epsilon$. Considering the time integrals in Eq.~\eqref{maseqnmkf5}, the validity of the foregoing limit is further justified through the numerical simulations described towards the end of this paper where $k$ and $|p-q|$ are found to peak around $\epsilon$.

Adopting the Born approximation \eqref{psikkppy} and keeping up to the first orders in $\nu$, we see that Eq.~\eqref{stij1kk} becomes
\begin{eqnarray}
\tau_{ij}(\kk,\pp,\pp')
&=&
\tau_{ij}^{(0)}(\kk,\pp,\pp')
+
\nu\,
\tau_{ij}^{(1)}(\kk,\pp,\pp')
\label{tkpp}
\end{eqnarray}
in terms of
\begin{eqnarray}
\tau_{ij}^{(0)}(\kk,\pp,\pp')
&=&
\frac{\hbar}{2}\,
\frac{P_{ijkl}(\kk) p_k p_l}{\sqrt{{p}{p'}}}\,
\delta(\pp-\pp'-\kk)
\,a_{\pp'}^\dag a_{\pp}
\label{tkpp0}
\pppp
\tau_{ij}^{(1)}(\kk,\pp,\pp')
&=&
\frac{\hbar}{2\pi^2}\,
\frac{P_{ijkl}(\kk)}
{\sqrt{{p}{p'}}[|\pp-\pp'-\kk|^2+\epsilon^2]}\,
\Big[
\frac{p_k p_l}{|\pp-\kk|^2-(p'-i\epsilon)^2}
+
\frac{p'_k p'_l}{|\pp'+\kk|^2-(p+i\epsilon)^2}
\Big]
a_{\pp'}^\dag a_{\pp}
\nppp&&
-(\epsilon\to\delta)
\label{tkpp1}
\end{eqnarray}
satisfying
\begin{eqnarray}
\tau_{ij}^{(0)\dag}(\kk,\pp,\pp')
&=&
\tau_{ij}^{(0)}(-\kk,\pp',\pp)
\label{tkpp0a}
\pppp
\tau_{ij}^{(1)\dag}(\kk,\pp,\pp')
&=&
\tau_{ij}^{(1)}(-\kk,\pp',\pp)
\label{tkpp1a}
\end{eqnarray}
which are consistent with Eq.~\eqref{aex}.

Substituting Eqs.~\eqref{tkpp} into Eq.~\eqref{maseqnmkf6} and repetitively using the same argument leading to free particles suffering no Markovian gravitational decoherence, we obtain the  2nd order (in $\nu$) asymptotic change of the density matrix
{\begin{eqnarray}
\Delta\rho
&=&
-
\frac{2\, G \nu^2}{\pi \hbar}
\int \dd^3 k\,\dd^3 p\,\dd^3 p'\,\dd^3 q\,\dd^3 q'\,
\frac{1}{k}
\,\delta(k+({q}-{q'}))
\,\delta(k-({p}-{p'}))
\nppp&&\hspace{8pt}
\;\;\times\;
[
\tau_{ij}^{(1)}(-\kk,\qq,\qq')
,\,
\tau_{ij}^{(1)}(\kk,\pp,\pp')
\rho_0
]
+ \hc
\label{maseqnmkf8}
\end{eqnarray}}

On account of Eqs.~\eqref{nu} and \eqref{tkpp1a}, Eq.~\eqref{maseqnmkf8} takes a more explicit form as follows
\begin{eqnarray}
\Delta\rho
&=&
-
\zeta
\int \dd^3 k\,\dd^3 p\,\dd^3 p'\,\dd^3 q\,\dd^3 q'\,
\,\delta(k-{p}+{p'})
\,\delta(k+{q}-{q'})
\frac{P_{ijkl}(\kk)}
{k\sqrt{{p}{p'}{q}{q'}}}
\nppp&&
\;\;\times\;
\Big\{
\frac{1}{[|\pp-\pp'-\kk|^2+\epsilon^2]}
\Big[
\frac{p_i p_j}{|\pp-\kk|^2-(p'-i\epsilon)^2}
+
\frac{p'_i p'_j}{|\pp'+\kk|^2-(p+i\epsilon)^2}
\Big]
-
(\epsilon\to\delta)
\Big\}
\nppp&&
\nppp&&
\;\;\times\;
\Big\{
\frac{1}{[|\qq-\qq'+\kk|^2+\epsilon^2]}
\Big[
\frac{q_k q_l}{|\qq+\kk|^2-(q'-i\epsilon)^2}
+
\frac{q'_k q'_l}{|\qq'-\kk|^2-(q+i\epsilon)^2}
\Big]
-
(\epsilon\to\delta)
\Big\}
\nppp&&
\;\;\times\;
\big[
a_{\qq'}^\dag a_{\qq}\,
,
a_{\pp'}^\dag a_{\pp}\,
\rho_0
\big]
+
\hc
\label{maseqnmkf11}
\end{eqnarray}
where
\begin{eqnarray}
\zeta
&=&
\frac{2\hbar\, G^3 M_\star^2 p_\ph^4}{\pi^5}
\label{zeta}
\end{eqnarray}
is a dimensionless parameter.

The action of Eq.~\eqref{maseqnmkf11} can be obtained from its action on an initial basis matrix element of the form
\begin{eqnarray}
\rho_0
&=&
\ket{\pp_0}\bra{\qq_0}
=
a_{\pp_0}^\dag
\ket{0}
\bra{0}a_{\qq_0} .
\label{rho0}
\end{eqnarray}

Substituting this form \eqref{rho0} into Eq.~\eqref{maseqnmkf11} and using the relation
\begin{eqnarray*}
\big[
a_{\qq'}^\dag a_{\qq}\,
,
a_{\pp'}^\dag a_{\pp}\,
\rho_0
\big]
&=&
\delta(\pp-\pp_0)
\delta(\qq-\pp')
\ket{\qq'}\bra{\qq_0}
-
\delta(\pp-\pp_0)
\delta(\qq'-\qq_0)
\ket{\pp'}\bra{\qq}
\end{eqnarray*}
obtained from Eq.~\eqref{comma}, we can usefully write the resulting $\Delta\rho$ as
\begin{eqnarray}
\Delta\rho
&=&
\Delta\varrho^\natural
+
\Delta\varrho^\flat
+
\hc
\label{delrho}
\end{eqnarray}
where
\begin{eqnarray}
\Delta\varrho^\natural
&=&
-\zeta
\int \dd^3 k\,\dd^3 p\,\dd^3 q\,
\,\delta(k+q-p_0)
\,\delta(k+q-p)
\frac{P_{ijkl}(\kk)\,\ket{\pp}\bra{\qq_0}}
{k q\sqrt{p p_0}}
\nppp&&
\;\;\times\;
\Big\{
\frac{1}{|\pp_0-\qq-\kk|^2+\epsilon^2}
\Big[
\frac{p_{0 i} p_{0 j}}{|\pp_0-\kk|^2-(q-i\epsilon)^2}
+
\frac{q_i q_j}{|\qq+\kk|^2-(p_0+i\epsilon)^2}
\Big]
-
(\epsilon\to\delta)
\Big\}
\nppp&&
\;\;\times\;
\Big\{
\frac{1}{|\qq-\pp+\kk|^2+\epsilon^2}
\Big[
\frac{q_k q_l}{|\qq+\kk|^2-(p-i\epsilon)^2}
+
\frac{p_k p_l}{|\pp-\kk|^2-(q+i\epsilon)^2}
\Big]
-
(\epsilon\to\delta)
\Big\}
\label{delrhon}
\end{eqnarray}
does not lose energy and
\begin{eqnarray}
\Delta\varrho^\flat
&=&
\zeta\!\!\!
\int\displaylimits_{\substack{k< \min \\ (p_0,q_0)}}\!\!\! \dd^3 k\,
\nppp&&
\hspace{-25pt}\times\;
P_{mnij}(\kk) \int\dd^3 p\,
\Big\{
\frac{\delta(k+p-p_0)}
{\sqrt{p_0 k p}\,
[|\pp_0-\pp-\kk|^2+\epsilon^2]}
\Big[
\frac{p_{0 i} p_{0 j}}{|\pp_0-\kk|^2-(p-i\epsilon)^2}
+
\frac{p_i p_j}{|\pp+\kk|^2-(p_0+i\epsilon)^2}
\Big]
-
(\epsilon\to\delta)
\Big\}
\ket{\pp}
\nppp&&
\hspace{-25pt}\times\;
P_{mnkl}(\kk) \int\dd^3 q\,
\Big\{
\frac{\delta(k+q-q_0)}
{\sqrt{q_0 k q}\,[|\qq_0-\qq-\kk|^2+\epsilon^2]}
\Big[
\frac{p_{0 k} p_{0 l}}{|\qq_0-\kk|^2+(q+i\epsilon)^2}
+
\frac{q_k q_l}{|\qq+\kk|^2-(q_0-i\epsilon)^2}
\Big]
-
(\epsilon\to\delta)
\Big\}
\bra{\qq}
\nppp
\label{dvrhm}
\end{eqnarray}
is responsible for dissipating photon energy by emitting bremsstrahlung gravitons.

\newpage

The setup above allows us to consider a wave packet profile
\begin{eqnarray}
\ket{\psi}
&=&
\int\dd^3 p\,
\psi(\pp)
\ket{\pp}
\label{psi0}
\end{eqnarray}
as an initial normalised state with
\begin{eqnarray}
\int\dd^3 p\,
|\psi(\pp)|^2
&=&
1
\label{psi0norm}
\end{eqnarray}
and a mean wave number vector $\pp_\ph$ so that
\begin{eqnarray}
\int\dd^3 p\,
\pp\, |\psi(\pp)|^2
&=&
\pp_\ph .
\label{psi0mean}
\end{eqnarray}

This can be used to construct an initial density matrix
\begin{eqnarray}
\rho_0
&=&
\ket{\psi}\bra{\psi}
=
\int\dd^3 p\int\dd^3 q\,
\psi(\pp)
\psi^*(\qq)\,
\ket{\pp}\bra{\qq} .
\label{rhopsi0}
\end{eqnarray}

Then from Eqs.~\eqref{rhopsi0} and \eqref{dvrhm}, we have
\begin{eqnarray}
\Delta\varrho^\flat
&=&
\zeta
\int \dd^3 k\,
\ket{p_{ij}}
\bra{p_{ij}}
=
\zeta
\int \dd^3 k\,
\big(
\ket{p_{+}}\bra{p_{+}}
+
\ket{p_{\times}}\bra{p_{\times}}
\big)
\label{Dvrhmpsi}
\end{eqnarray}
where
\begin{eqnarray}
\ket{p_{ij}}
&=&
\ket{p_{ij}}(\kk)
\nppp
&=&
P_{ijkl}(\kk)\int\dd^3 p'\int\dd^3 p\,
\psi(\pp')\,
\delta(k+p-p')\,
\frac{1}{\sqrt{p' k p}}
\nppp&&
\times\;
\Big\{
\frac{1}
{|\pp'-\pp-\kk|^2+\epsilon^2}
\Big[
\frac{p'_{k} p'_{l}}{|\pp'-\kk|^2-(p-i\epsilon)^2}
+
\frac{p_k p_l}{|\pp+\kk|^2-(p'+i\epsilon)^2}
\Big]
-
(\epsilon\to\delta)
\Big\}
\ket{\pp}
\nppp
&=&
\frac{1}{p_\ph^3}\int\displaylimits_{p'>k}\dd^3 p'
\int\displaylimits_{p=p'-k}\dd\Omega_p\,
A_{ij}\ket{\pp}
\label{kpijpsi}
\end{eqnarray}
with the dimensionless TT amplitude
\begin{eqnarray*}
A_{ij}
&=&
A_{ij}(\kk,\pp',\Omega_p)
\nppp
&:=&
P_{ijkl}(\kk)\,\psi(\pp')\,
\frac{p_\ph^3 p^{3/2}}
{\sqrt{p' k}}
\nppp&&
\times\;
\Big\{
\frac{1}
{|\pp'-\pp-\kk|^2+\epsilon^2}
\Big[
\frac{p'_{k} p'_{l}}{|\pp'-\kk|^2-(p-i\epsilon)^2}
+
\frac{p_k p_l}{|\pp+\kk|^2-(p'+i\epsilon)^2}
\Big]
-
(\epsilon\to\delta)
\Big\}
\Big|_{p=p'-k} .
\end{eqnarray*}

\section{Gravitational Bremsstrahlung with a Single Momentum Initial State}
\label{sec.singlemom}

For simplicity, we now restrict to single momentum initial state, deferring the more general wave-packet initial states to a future investigation.

Such a state is obtained by setting $\qq_0 = \pp_0 = \pp_\ph$ in Eq.~\eqref{rho0} so that Eq.~\eqref{dvrhm} takes the form
\begin{eqnarray}
\Delta\varrho^\flat
&=&
\zeta
\int_{k < p_\ph} \dd^3 k\,
\ket{p_{ij}}
\bra{p_{ij}}
=
\zeta
\int_{k < p_\ph} \dd^3 k\,
\big(
\ket{p_{+}}\bra{p_{+}}
+
\ket{p_{\times}}\bra{p_{\times}}
\big)
\label{Dvrhm}
\end{eqnarray}
where
\begin{eqnarray}
\ket{p_{ij}}
&=&
\ket{p_{ij}}(\kk)
=
p_\ph^{-3/2}
\int\displaylimits_{p=p_\ph-k}\dd\Omega_p\,
A_{ij}\ket{\pp}
\label{kpij}
\end{eqnarray}
with the dimensionless TT amplitude
\begin{eqnarray*}
A_{ij}
&=&
A_{ij}(\kk,\Omega_p)
\nppp
&:=&
\frac{p_\ph p^{3/2}P_{ijkl}(\kk)\,}{\sqrt{k}}
\Big\{
\frac{1}{\pp_\ph-\pp-\kk|^2+\epsilon^2}
\Big[
\frac{p_{0 k} p_{0 l}}{|\pp_\ph-\kk|^2-(p-i\epsilon)^2}
+
\frac{p_k p_l}{|\pp+\kk|^2-(p_\ph+i\epsilon)^2}
\Big]
-
(\epsilon\to\delta)
\Big\}
\Big|_{p=p_\ph-k} .
\end{eqnarray*}

There are two orthogonal parts of $A_{ij}$:
\begin{eqnarray*}
A_{ij}
&=&
A_{ij}^{+} + A_{ij}^{\times}
\end{eqnarray*}
corresponding to the two ($+$ and $\times$) polarisations of the gravitational waves \cite{Flanagan2005},
so that the gravitational wave square amplitude decomposes accordingly as
\begin{eqnarray}
|A|^2
&:=&
A_{ij}A_{ij}^*
=
A_{ij}^{+}A_{ij}^{+ *} + A_{ij}^{\times}A_{ij}^{\times *} .
\label{Asq0}
\end{eqnarray}

For numerical evaluations, we can conveniently choose $\pp_\ph = p_\ph \hat{\zz}$, then we have
\begin{eqnarray*}
A_{ij}
&=&
\frac{p_\ph p^{3/2}}{\sqrt{k}}
\Big\{
\frac{1}{\pp_\ph-\pp-\kk|^2+\epsilon^2}
\Big[
\frac{p_\ph^2 \delta_{3 k} \delta_{3 l}}{|\pp_\ph-\kk|^2-(p-i\epsilon)^2}
+
\frac{p_k p_l}{|\pp + \kk|^2-(p_\ph+i\epsilon)^2}
\Big]
-
(\epsilon\to\delta)
\Big\}
\Big|_{p=p_\ph-k} .
\end{eqnarray*}

For example, in the limit of a point source of gravity with effective radius $1/\delta \to 0$, Eq.~\eqref{Asq0} yields
\begin{eqnarray}
|A|^2_\epsilon
&=&
\frac{p_\ph^2 p^3}
{k\,[|\pp_\ph-\pp-\kk|^2+\epsilon^2]^2}
\bigg\{
\frac{p_\ph^4 P_{3333}(\kk)}
{\big||\pp_\ph-\kk|^2-(p-i\epsilon)^2\big|^2}
\nppp&&
+
\frac{2\,p_\ph^2 p^2 P_{33ij}(\kk) \hat{p}_i \hat{p}_j}
{\Re\big[(|\pp_\ph-\kk|^2-(p+i\epsilon)^2)(|\pp+\kk|^2-(p_\ph+i\epsilon)^2)\big]}
+
\frac{p^4 P_{ijkl}(\kk)\,\hat{p}_i \hat{p}_j \hat{p}_k \hat{p}_l}
{\big||\pp+\kk|^2-(p_\ph+i\epsilon)^2\big|^2}
\bigg\}
\bigg|_{p=p_\ph-k}
\label{Asq}
\end{eqnarray}
where
\begin{eqnarray*}
\big||\pp_\ph-\kk|^2-(p-i\epsilon)^2\big|^2
&=&
(|\pp_\ph-\kk|^2-p^2
+
\epsilon^2)^2
+
4\epsilon^2 p^2 ,
\nppp
\Re\big[(|\pp_\ph-\kk|^2-(p+i\epsilon)^2)(|\pp+\kk|^2-(p_\ph+i\epsilon)^2)\big]
&=&
(|\pp_\ph-\kk|^2 - p^2 + \epsilon^2)
(|\pp+\kk|^2 - p_\ph^2 + \epsilon^2)
-
4 \epsilon^2 p_\ph p ,
\nppp
\big||\pp+\kk|^2-(p_\ph+i\epsilon)^2\big|^2
&=&
(|\pp+\kk|^2-p_\ph^2
+
\epsilon^2)^2
+
4\epsilon^2 p_\ph^2 ,
\end{eqnarray*}
and
\begin{eqnarray*}
P_{3333}(\kk)
&=&
\frac12(1 - \hat{k}_3^2)^2 ,
\nppp
P_{33ij}(\kk) \hat{p}_i \hat{p}_j
&=&
\hat{p}_3^2
-
2 (\hat{k}_i \hat{p}_i) \hat{k}_3 \hat{p}_3
+
\frac{1}{2}
\big[
(\hat{k}_i \hat{p}_i)^2 (\hat{k}_3^2 + 1)
+
\hat{k}_3^2
-
1
\big] ,
\nppp
P_{ijkl}(\kk) \hat{p}_i \hat{p}_j \hat{p}_k \hat{p}_l
&=&
\big[
1 - (\hat{k}_i \hat{p}_i)^2
\big]^2 .
\end{eqnarray*}

It is also useful to introduce $|A|^2_\delta := |A|^2_{\epsilon\to\delta}$.
For simplicity, let us adopt as a reasonable first order-of-magnitude estimate of the gravitational wave square amplitude for a source of gravity with an effective range $1/\epsilon$ and radius $1/\delta$ to be
\begin{eqnarray}
|A|^2_{\epsilon,\delta} := |A|^2_\epsilon - |A|^2_\delta .
\label{A2est}
\end{eqnarray}

The total dissipated outgoing energy corresponding to Eqs.~\eqref{delrho} and \eqref{Dvrhm} then follows as
\begin{eqnarray}
\Delta E
&=& 
\frac{2\zeta}{p_\ph^3}
\int\dd^3 k\,
\int\dd\Omega_p\,\hbar k\,|A|^2_{\epsilon,\delta}(\kk,\Omega_p)
\nppp
&=& 
\frac{2\zeta\hbar}{p_\ph^3}
\int_0^{p_\ph} \dd k\,k^3
\int\dd\Omega_k
\int\dd\Omega_p\,|A|^2_{\epsilon,\delta}(\kk,\Omega_p) .
\label{DelE}
\end{eqnarray}

\newpage

To obtain an expression in term of dimensionless quantities, we express $\pp_\ph,\pp,\kk,\epsilon$ as dimensionless quantities in units of $p_\ph$.

Then, by replacing $\pp_\ph \to p_\ph\pp_\ph,\pp \to p_\ph\pp,\kk \to p_\ph\kk,\epsilon \to p_\ph\epsilon, \delta \to p_\ph\delta$ and using Eq.~\eqref{zeta} with the mass-energy of the gravity source $E_\star=M_\star$, we arrive at the fractional energy loss $\mu:=\Delta E/E_\ph=\Delta E/(\hbar p_\ph)$ given by
\begin{eqnarray}
\mu
&=&
\frac{4}{\pi^5}\,\hbar\, G^3 E_\star^2 p_\ph^4\,
\Big(
\frac{S_\epsilon}{\epsilon}
-
\frac{S_\delta}{\delta}
\Big)
\nppp
&=&
\frac{4}{\pi^5}\,\frac{E_\star^2 E_\ph^4}{\EP^{\,6}}\,
\Big(
\frac{S_\epsilon}{\epsilon}
-
\frac{S_\delta}{\delta}
\Big)
\label{loss}
\end{eqnarray}
where $\EP$ is the Planck energy and
\begin{eqnarray}
S_\epsilon
&=&
\epsilon
\int_0^{1} \dd k\,k^3
\int\dd\Omega_k
\int\dd\Omega_p\,|A|^2_\epsilon(k,\Omega_k,\Omega_p)
\nppp
&=&
\int_0^{1} \dd k\,
\int\dd\theta_k\int\dd\phi_k
\int\dd\theta_p\int\dd\phi_p\,
F_\epsilon(k,\theta_k,\phi_k,\theta_p,\phi_p)
\label{DelEK}
\end{eqnarray}
with $|A|^2_\epsilon$ obtained from Eq.~\eqref{Asq} for $p_\ph\to1$ and
\begin{eqnarray}
\lefteqn{
F_\epsilon(k,\theta_k,\phi_k,\theta_p,\phi_p)
}
\nppp
&=&
\frac{\epsilon k^2 p^3 \sin\theta_k \sin\theta_p}
{[|\pp_\ph-\pp-\kk|^2+\epsilon^2]^2}
\bigg\{
\frac{\frac12(1 - \hat{k}_3^2)^2}
{(|\pp_\ph-\kk|^2-p^2)^2
+
2\epsilon^2 (|\pp_\ph-\kk|^2+p^2)
+
\epsilon^4}
\nppp&&
+
\frac{2\,p^2
\Big[
\hat{p}_3^2
-
2 (\hat{k}_i \hat{p}_i) \hat{k}_3 \hat{p}_3
+
\frac{1}{2}
\big[
(\hat{k}_i \hat{p}_i)^2 (\hat{k}_3^2 + 1)
+
\hat{k}_3^2
-
1
\big]
\Big]
}
{(|\pp_\ph-\kk|^2 - p^2 + \epsilon^2)
(|\pp+\kk|^2 - 1 + \epsilon^2)
-
4 \epsilon^2 p}
+
\frac{p^4 \big[
1 - (\hat{k}_i \hat{p}_i)^2
\big]^2}
{(|\pp+\kk|^2-1)^2
+
2\epsilon^2 (|\pp+\kk|^2+1)
+
\epsilon^4}
\bigg\}
\bigg|_{p=1-k} .
\nppp
\label{Asq2}
\end{eqnarray}

We also have analogous constructions for $S_\delta$ and $F_\delta$.
Numerical evaluations of
$F_\epsilon$ and $F_\delta$ as functions of $(k,\theta_k,\phi_k,\theta_p,\phi_p)$
show sharp but finite peaks around small $k$, $\theta_k$ and $\theta_p$ for small $\epsilon, \delta$. This leads to finite numerical integrations with
$S_\epsilon\approx S_\delta\approx 100$ for $\epsilon, \delta \ll 1$.
Therefore, for a gravity source with a characteristic radius $1/\delta$ and effective range $1/\epsilon$ in units of $1/p_\ph$, from Eq.~\eqref{loss} we have
\begin{eqnarray}
\mu
&=&
\frac{E_\star^2 E_\ph^4}{\EP^{\,6}}\,
\Big(\frac{1}{\epsilon}-\frac{1}{\delta}\Big) .
\label{totloss}
\end{eqnarray}
approximately, corresponding to the rough spread of the photon impact parameter to be from the surface of the gravitational lens to one radius from the surface.

The energy loss rate above can be interpreted $\mu \approx \Gamma\; \tau$ where
$\Gamma = {E_\star^2 E_\ph^4}/{\EP^{\,6}}$
is the effective graviton emission transition rate by the photon, and
$\tau={1}/{\epsilon}-{1}/{\delta}$
is the effective interaction time between the photon and the lensing mass $M_\star$ with the corresponding effective interaction range $R_\rng=1/\epsilon$. See Fig.~\ref{fig.1}.
This is consistent with the long travel time or distance assumption stated in section \ref{sec:cgq}.

In full physical units, Eq.~\eqref{totloss} becomes
\begin{eqnarray}
\mu
&=&
{\frac{\hbar\, G^3}{c^{12}}\,
\Big(R_\rng-R_\star\Big)M_\star^2 \omega_\ph^5}
\label{totloss2}
\end{eqnarray}
where
the speed of light $c$ has now been reinstated.

\newpage

\section{Conclusions and Discussion}

Based on the gravitational quantum vacuum, which has recently been shown to lead to gravitational decoherence \cite{Oniga2016a} and gravitational spontaneous radiation that recovers the well-established quadrupole radiation \cite{Oniga2017b}, in this paper we have presented, to our knowledge, the first approach to the spontaneous bremsstrahlung of light due to the combined effects of gravitational lensing and spacetime fluctuations. Our present work yields a new quantum gravitational mechanism whereby starlight emits soft gravitons and becomes partially redshifted. This effect may contribute to the stochastic gravitational wave background \cite{Quinones2017, Christensen2019, Allen1996}. We also note that while the term \eqref{delrhon} for the outgoing light does not undergo photon to graviton energy conversion, it exhibits a type of recoherence of photons \cite{Bouchard2015}.

\begin{figure}
\begin{center}
\includegraphics[width=0.8\linewidth]{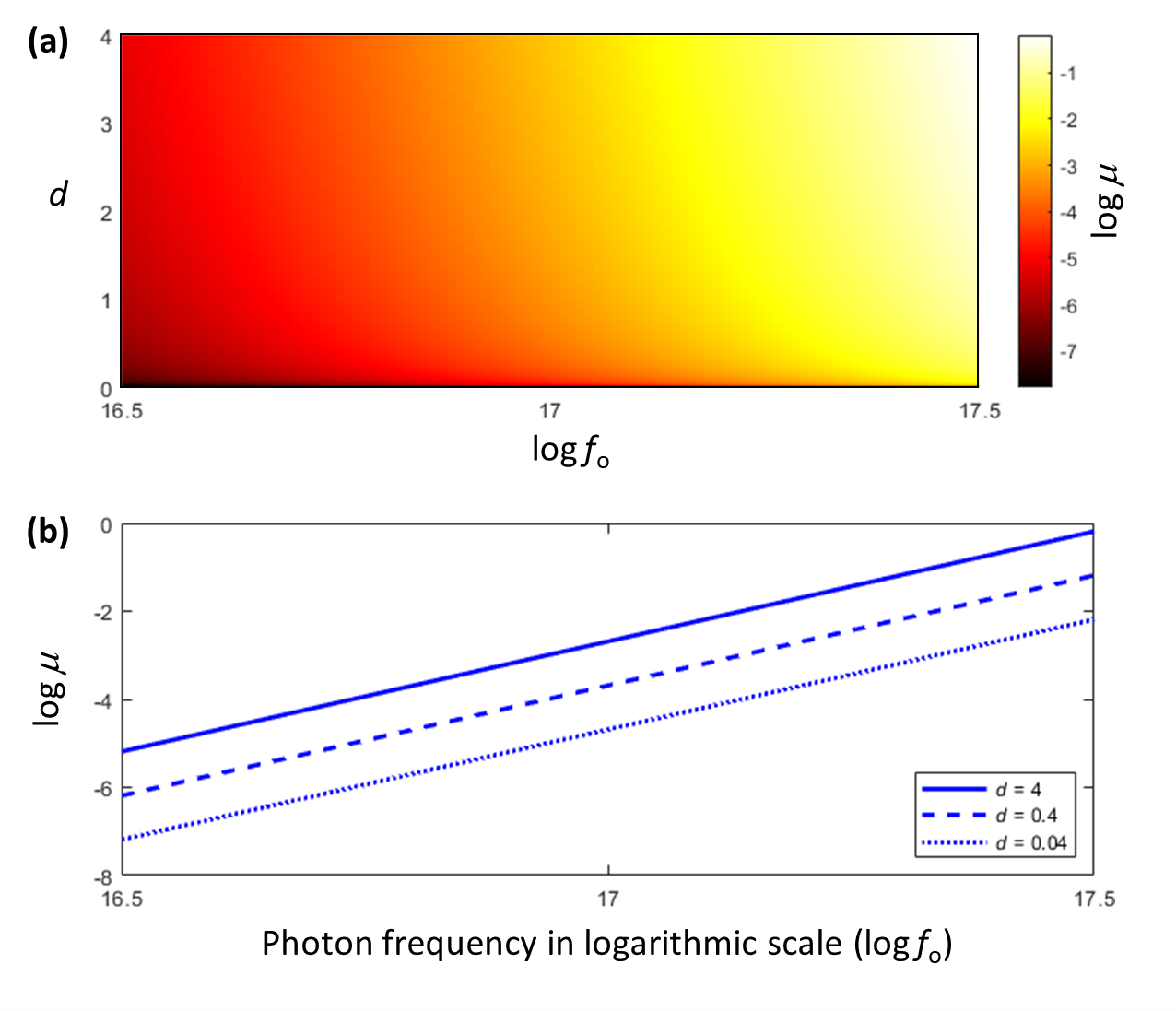}
\caption{The fractional energy loss $\mu$ given by Eq.~\eqref{totloss2} as a function of the photon frequency $f_\ph$ (Hz) in logarithmic scale along the horizontal axis evaluated with a chosen range of the effective gravity range parameter $d=(R_\rng-R_\star)/R_\star$ for the X-ray binary Cygnus X-1. Above, $\log \mu$ is plotted in (a) as a projected height onto the $(\log f_\ph, d)$ plane and in (b) as a line for 3 representative values of $d$.
It is obvious that as the photon frequency $f_\ph$ increases from the optical spectrum, the photon-to-graviton energy conversion ``chips in'' at around $f_\ph = 3\times10^{17}$ Hz at the start of the soft X-ray spectrum, so that $\mu \sim 0.01$ for $d=4$ corresponding to $R_\rng \sim 100 R_\odot$. Although here $\mu$ appears to carry on increasing with $f_\ph$, the validity of Eq.~\eqref{totloss2} is strictly limited to small $\mu$ due to the Born approximation with weak interactions we have assumed in Eq.~\eqref{psikkpp} used to derive $\mu$ in Eq.~\eqref{totloss2}.
Nonetheless, the ascending trend of $\mu$ with the photon frequency $f_\ph$ suggests that starting from the soft X-ray frequencies, the gravitational bremsstrahlung of light in the vicinity of Cygnus X-1 may be important.}
\label{fig.2}
\end{center}
\end{figure}

\begin{figure}
\begin{center}
\includegraphics[width=0.8\linewidth]{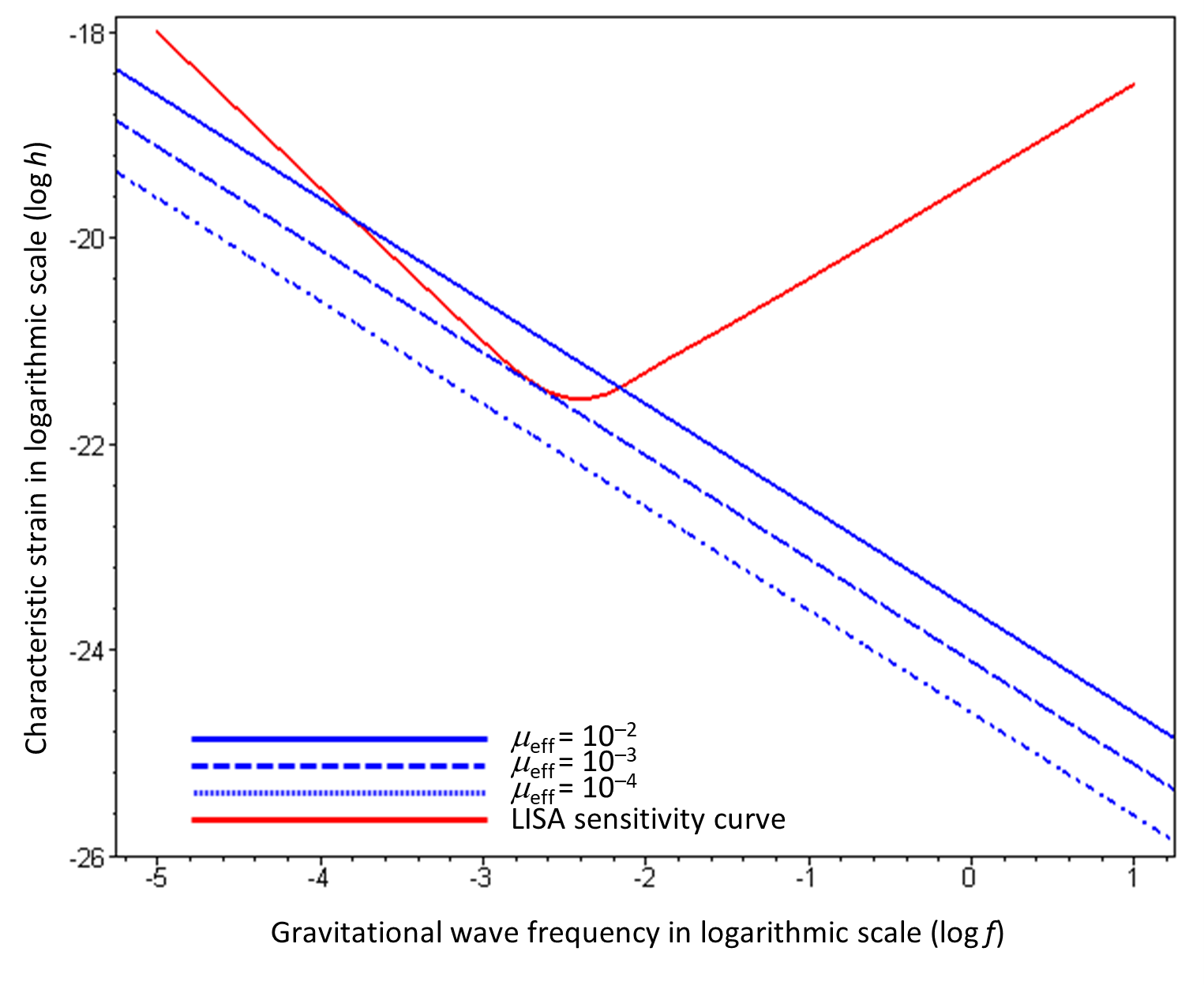}
\caption{Estimated characteristic strain $h$ of the gravitational waves with effective photon-to-graviton energy transfer rate $\mu_\eff$ for Cygnus X-1 with $0.01\% \le \mu_\eff \le 1\%$ against the gravitational wave frequency $f$ in Hz compared with the LISA characteristic strain sensitivity curve \cite{Moore2015}.}
\label{fig.3}
\end{center}
\end{figure}

Our work naturally raises the prospect of potential detection of the released stochastic gravitational waves. Addressing this question in detail requires a further investigation which is currently underway by the authors. It is however of interest at the present stage to envisage a plausible observation scenario. To this end, let us take the well studied strong X-ray binary Cygnus X-1 \cite{Bel2006} because this system has both a large gravitational field and a strong X-ray source {\red similar to that illustrated in Fig.~\ref{fig.1}. However, both the compact object/black hole and the companion supergiant star in the Cygnus X-1 binary system are massive, with a total mass $M_\star \approx 40 M_\odot$ and an orbiting radius $\approx 20 R_\odot$, which we will take as the effective $R_\star$. This} would make $R_\rng$ to be in the region of $100 R_\odot$ and so from the discussions in Sec.~\ref{sec.singlemom}, the spontaneously emitted stochastic gravitational waves would have a mean wavelength in the same region having a mean frequency $f \sim 0.01$ Hz. Using these parameters, the fractional energy loss $\mu$ in Eq.~\eqref{totloss2} {\red is plotted in Fig.~\ref{fig.2} with the effective gravity range parameter $d:= (R_\rng-R_\star)/R_\star $ chosen to be $0 \le d \le 4$, corresponding to the rough spread of the photon impact parameter to be from the surface of the gravitational lens to $R_\rng \approx 100 R_\odot$ discussed above.} The detection of sub-Hz gravitational waves is a unique strength of the proposed LISA mission, which is expected to reach a corresponding characteristic strain sensitivity close to $h \sim 10^{-22}$ \cite{Moore2015}, as shown in Fig.~\ref{fig.3}.

From the X-ray luminosity of Cygnus X-1 $L_X\approx 4 \times 10^{37}$ erg s$^{-1}$ \cite{Liang1984, Wen1999} and its distance $D\approx 6100$ light {\red years to the Earth, one gets the arriving X-ray energy flux to be} $S_{X} = L_X/(4\pi D^2)\approx 10^{-7}$ {erg cm$^{-2}$s$^{-1}$}.
Let us suppose that the total gravitational wave luminosity of Cygnus X-1 arises from $L_\text{GW}=\mu_\eff L_X$, in terms of an effective photon-to-graviton energy transfer rate $0< \mu_\eff < 1$. We can then estimate the characteristic strain $h$ of the gravitational waves with energy density expression $U_\text{GW} = c^2 \omega^2 h^2/(32 \pi G)$ where $h^2 = h_+^2 + h_\times^2$ with $\omega=2\pi f$ and the energy flux $S_\text{GW} = c\, U_\text{GW}$ so that $S_\text{GW} =  \mu_\eff S_X$. For example, if $ \mu_\eff = 0.1\%$ and $f=0.01$ Hz, then $S_\text{GW} \approx 10^{-10}$ {erg cm$^{-2}$s$^{-1}$} yielding $h \approx 10^{-22}$.

As shown in Fig.~\ref{fig.3}, we choose moderate effective transfer rate values $0.01\% \le \mu_\eff \le 1\%$ across the gravitational wave frequencies $10^{-5}$ Hz $\le f \le$ 10 Hz. Indeed, for $\mu_\eff \gtrsim 10^{-3}$ at around $f \sim$ 0.01 Hz, the characteristic strain of the bremsstrahlung gravitational waves from Cygnus X-1 could be above the sensitivity level of LISA and hence potentially detectable. Furthermore, the buildup of similar sources could also contribute to an overall stochastic gravitational wave background. Based on the initial results and estimates reported in this work, we plan to analyse and quantify the properties of the bremsstrahlung gravitational waves from intense astronomical sources of light and gravity in a more realistic setting for future publication.

\section*{Acknowledgements}
C.W. is indebted to John S. Reid for stimulating discussions and grateful to the Cruickshank Trust for financial support.  M.M. would like to thank the Student Awards Agency Scotland (SAAS) for funding.


\end{document}